\documentclass[aps,showpacs,twocolumn]{revtex4}

\usepackage{graphicx}

\usepackage{dcolumn}
\usepackage{color}

\begin{document}

\title{ BEC-BCS crossover of a trapped Fermi gas without using the local density
approximation}

\author{R. J\'auregui, R. Paredes, and G. Toledo
S\'anchez} \affiliation{ Instituto de F\'{\i}sica, Universidad
Nacional Aut\'onoma de M\'exico, Apdo. Postal 20-364, M\'exico D.
F. 01000, M\'exico.}

\date{\today}


\begin{abstract}
We  perform a variational quantum Monte Carlo simulation of an
interacting Fermi gas confined in a three dimensional harmonic
potential. This gas is considered as the precursor system from which
a molecular bosonic gas is formed. Based on the results of two-body
calculations for trapped atoms, we propose a family of variational
many-body wave functions that takes into account the qualitative
different nature of the BCS and BEC regimes as a function of the
scattering length. Energies, densities and correlation functions are
calculated and compared with previous results for homogeneous gases.
Universality tests at the unitarity limit are performed including
the verification of the virial relation and the evaluation of the
$\beta$ parameter.

\end{abstract}


\maketitle

The crossover from a low interacting attractive
Bardeen-Cooper-Schieffer (BCS) gas to a molecular Bose-Einstein
condensate (BEC) has been realized experimentally using a mixture of
ultracold Fermi atoms in two hyperfine states
\cite{experiments,thomas,bartenstein,salomon,bourdel}. In dilute
Fermi gases, the atomic interactions have a range much smaller than
the interparticle separation. Nevertheless, under proper conditions,
a magnetic field can be used to tune the attractive potential so
that the energy of a pair of scattering atoms is close to that  of a
molecular bound state (Feshbach resonance). In experiments where the
resonance is broad, it can be represented by a single-channel model
in which  the $s$-wave scattering length $a$ determines the general
features of the ultracold atomic gas. At low enough temperatures,
atoms in different hyperfine states pair into bound molecules for
positive values of $a$, forming a molecular BEC that can be
adiabatically converted into a degenerate Fermi gas by shifting $a$
to negative values. At resonance ($\vert a \vert \rightarrow \infty$),
the gas acquires universal properties, i. e., they
are independent of any feature of the  atomic potentials
\cite{thomas,heiselberg, ho}. This is the so called unitarity limit.

The theoretical description of the BEC-BCS crossover is
complicated because there is no {\it ad hoc} single parameter to
control the relevant features in both regimes. A reasonable
alternative to infer properties of the system at unitarity has
been to consider an $homogeneous$  Fermi gas, and to assume the
universality hypothesis according to which the only dominant
length is the interparticle separation. Then, the thermodynamic
potentials have a universal form specified by only few universal
numbers \cite{ho}. For instance, the interaction energy is proportional to the
Fermi energy via a universal constant $\beta$.

Quantum Monte Carlo (QMC) techniques support  $ab$ $initio$ methods
to theoretically test the universality hypothesis. They can be used
to approximately solve the many-body Schr\"odinger equation for a
given  model of the interaction potential. In previous studies the
BEC and BCS regimes have been explored using a fixed-node QMC
technique,  which is rigorous but also  computationally demanding.
In particular, the value of $\beta$ has been predicted
 considering up to 66 particles \cite{carlson,giorgini}.
Nevertheless, these QMC calculations have been performed for
homogeneous gases despite the fact that experimentally the
atoms are confined by an external field and therefore the
inhomogeneity is intrinsic to the
problem. Comparison of QMC results with experiments are then based
on a mapping of the trapped system onto a corresponding
homogeneous problem with a local value of the Fermi energy
$\epsilon_F(x)$ (local density approximation). This reasoning also
allows to study the universality hypothesis consequences for a
trapped dilute Fermi gas in the crossover \cite{thomas, kinast,
virial}.

Here we study the behavior of a balanced mixture of $2N$ interacting
trapped Fermi atoms, using a variational quantum Monte Carlo (VQMC)
technique, which allows to deal with large number of particles but
strongly depends on the choice of the variational wave function. We
compute the many-body ground state of the inhomogeneous gas from the
BCS to the BEC regime for a given short range interaction potential.
For simplicity we shall consider an isotropic harmonic potential.
The values of $N$ are the highest reported in QMC calculations for
this kind of atomic systems. Emphasis will be made on the energies,
densities and two point correlation functions. At the unitarity
limit,  $\beta$ will be directly evaluated and the validity of the
virial theorem verified \cite{virial}.

To describe the many-body system, let us first consider the
attractive two-body interaction potential $V(r_{ij})=-
V_0e^{-2\vert \vec r_i^\uparrow - \vec r_j^\downarrow\vert/b}$ of
range $b/2$, to model the effective interaction between atoms in
different hyperfine states. The corresponding two-body problem for
a particle of reduced mass $m/2$, $[{\hat p^2}/m  + V]\varphi =
E\varphi$, has analytical $s$-wave solutions \cite{rarita}
$\varphi(r_{ij}) = u(r_{ij})/r_{ij}$ both in the continuum ($ u(y)
= c_1J_{i b\sqrt{E m}/\hbar}(y)+c_2J_{-ib\sqrt{E m}/\hbar}(y)$)
and in the bound region ($u(y) = c_+J_{ b\sqrt{\vert E\vert
m}/\hbar}(y)$). Here $y = (b\sqrt{V_0 m}/\hbar)e^{-r/b}$ and
$J_\nu$ represents the  first kind Bessel function of order
$\nu$. The scattering length is
\begin{equation}
a = -b\Big[ \frac{\pi}{2} \frac{N_0(b\sqrt{V_0 m}/\hbar)}
{J_0(b\sqrt{V_0 m}/\hbar)} - \log\big( b\sqrt{V_0m}/2\hbar\big) -C
\Big]
\end{equation}
with $N_0$ the second kind Bessel function of zero order, and $C$
the Euler constant. This scattering length diverges whenever
$J_0(b\sqrt{V_0m}/\hbar)=0$. If  $z_n, n=0,1,2,...$ are the zeros of
this Bessel function in increasing order, the potential $V(r_{ij})$
admits just $n$-bound states for $z_n< b\sqrt{V_0 m}/\hbar <
z_{n+1}$. The discrete eigenvalues are determined by the boundary
condition at $r_{ij}=0$, $J_{b\sqrt{\vert E\vert
m}/\hbar}(b\sqrt{V_0m}/\hbar)=0$. Taking into account the presence
of a spherical harmonic trap the Schr\"odinger equation, $[{\hat
p^2}/{m } + m \omega^2r^2/4+ V(r)]\varphi(r) = E\varphi(r)$, can be
numerically solved and some interesting results are obtained when
the range of the potential $b/2<< \sqrt{\hbar/m\omega}$
\cite{becbcs}. For instance: (i) The ground state energy $E_0$ for
$b\sqrt{V_0 m}/\hbar =z_0$ (which implies $\vert
a\vert\rightarrow\infty$ and a zero-energy resonance in the
homogeneous problem) becomes $E_0\sim 0.5\hbar\omega$ as $b\sim 0$
(e.g. for $b=0.03 \sqrt{\hbar/m\omega}$,
$E_0=0.510655588\hbar\omega$) while the $s$-wave excited states have
energies $\sim (2n_e +0.5)\hbar\omega$; (ii) the ground state for
$z_0\le b\sqrt{V_0 m}/\hbar<z_1$ can be well approximated by
$u(y(r)){\rm exp}(-mr^2/2\hbar\omega)/r$. Result (i) is consistent
with the analytical solution of the two-body problem in the presence
of a contact interaction with a coupling constant determined by the
scattering length \cite{wilkens}.

 To address the many-body system we rely on the
VQMC method. A trial wave function $\phi_T$ is assumed and, from
initial stochastically generated events, a Metropolis Monte Carlo
algorithm samples the distribution
$\vert\phi_T\vert^2/\int\vert\phi_T\vert^2$ for variations of the
atoms positions. After a thermalization process, the  energy is
evaluated at each step. The average energy converges to its
expectation value, provided enough points are taken to sample.
 The parameters used to characterize $\phi_T$ are then varied to search
a minimum of the energy. The strength of this method relies on the
proper choice of the trial wave function. Here, the trial wave
function in the region of negative scattering length,
$0<b\sqrt{V_0 m}/\hbar<z_0$, has the Jastrow-Slater form
\begin{equation}
\Psi_{\lambda}(x)=e^{-\lambda  \sum_{i,j=1}^N V(r_{ij})}
\Phi_{FG}(\sqrt{\omega^\prime/\omega}x) \label{Psivar}
\end{equation}
where $\lambda$ and $\omega^\prime$ are variational parameters,
and $\Phi_{FG}(x)$ is the Fermi gas wave function given by a
product of Slater determinants (one for each hyperfine state)
describing a noninteracting system of harmonically trapped atoms.
 This variational wave function has the advantage of being exact when no
interactions between hyperfine states are allowed ($\lambda=0$,
$\omega^\prime=\omega$ ) and does not require an explicit
introduction of a healing distance \cite{carlson,healing}. It is inspired on previous
calculations for the nuclear matter \cite{nuclear}, where an
appropriate choice of the potential allows to explore dynamically
the interplay of the nuclear-to-quark matter regime while being
exact in both limits. Other forms of the Jastrow wave function can
be found in Refs.~\cite{carlson,giorgini}.

In the region of non-negative scattering length and $z_0 \le
b\sqrt{V_0 m}/\hbar<z_1$, the Jastrow-Slater wave function over
estimates the energy. Therefore, we propose the following trial wave
function
\begin{equation}
\Psi_\lambda (x) = {\cal
A}\phi_\lambda(1,1^\prime)...\phi_\lambda(N,N^\prime)\label{eq:wfa}
\end{equation}
where ${\cal A}$ is the antisymmetrizer operator and
\begin{equation}
\phi_\lambda(i,j) = u(y(r_{ij}))e^{(-m\vert \vec
r_{ij}\vert^2/4\hbar\omega)} e^{(-\lambda m\vert \vec R_{ij}\vert
^2\hbar\omega)}/r_{ij} \label{eq:wfb}\end{equation} with $\vec
r_{ij}$ and $\vec R_{ij}$ the relative and center of mass
coordinates associated to $\vec r_i^\uparrow$ and $\vec
r_j^\downarrow$; $\lambda$ is the variational parameter. The
structure of this trial wave function is a variational extension
for the inhomogeneous gas from that proposed in
Ref.~\cite{giorgini}.

 Given $2N$ particles, a fixed value of $b$
and a scattering length $a$, we determine the energy for a set of
values of the variational parameters to pick up the optimal. Each
run used about $10^3$ steps for thermalization and about $10^4$
more to take data. In addition, we estimate the dependence on the
initial conditions that might be not erased in the thermalization
process. The quoted errors take into account all the above
factors.

The results obtained for the optimal energy per particle as a
function of the scattering length $a$ are shown in Fig.~1. The
values are normalized as described
 in its caption.  By construction these normalized
energies take the asymptotic value of  zero (one) for small
positive (negative) values of $a$. The fact that
these energies fall in a curve almost independent of the value of
$N$ reflects that the Fermi wave number $k_F$, associated to the
ideal Fermi gas energy $E_{IFG}=\hbar^2 k_F^2/2m \approx
(6N)^{1/3}\hbar \omega $, defines the proper distance scale to
measure  $a$. It also means that the same curve
can be expected for larger values of $N$ and supports a universal
behavior in the crossover.

Notice that in the BCS side of the crossover the interaction
energy is very small compared to $E_{IFG}$, in accordance with
previous  QMC calculations in  homogeneous gases
\cite{carlson,giorgini}. In fact, we have checked that for $-k_F a
<1$ this variational energy coincides with the result obtained
using an effective contact interaction \cite{us}. At unitarity and
for $a>0$, the role of the interaction energy becomes more
relevant. It is found that in the extreme BEC limit
$k_Fa\rightarrow 0^+$ there is a strong competition between Pauli
blocking effects and the attractive interaction between the atoms
in different hyperfine states. This leads to numerical precision
problems that could indicate that this highly correlated system
should be described beyond the simple scheme of almost
non-interacting molecules.

The $\beta$ parameter for a trapped gas is usually evaluated
according to the following reasoning \cite{thomas, kinast}. Within
the universality hypothesis, for a trapped gas at unitarity,
the equation of state would be $(1+\beta)\epsilon_F +
U_{HO} = \mu_0$, with $\epsilon_F(x)$ the $local$ Fermi energy,
$U_{HO}$ the trapping potential, and $\mu_0$ the $global$ chemical
potential. This equation is equivalent to that of a trapped
noninteracting gas of particles with an effective mass
$m/(1+\beta)$ \cite{baker} so that the effective chemical
potential is also simply scaled to $\mu_0 =
E_{IFG}\sqrt{1+\beta}$. The total energy at $T=0$, which
determines the energy scale is then:
\begin{equation}
E_U=E_{IFG} \sqrt{1+\beta}. \label{beta}
\end{equation}
Since we are working with few atoms it is necessary to take into
account the exact expression for $E_{IFG}$. For a closed shell
configuration (all single particle states with energies below
$E_F=({\cal M}_F + 3/2)\hbar\omega$ are occupied) this energy is
given by \cite{us}
\begin{equation}
E_{IFG}/2N\hbar\omega= (3/4)({\cal M}_F + 2), \label{dtfg}
\end{equation}
instead of the large $N$ limit, $E_{IFG}/2N=3E_F/4$,  which
produces a slightly underestimated value.

 The universality
hypothesis can be tested independently through a virial relation
\cite{virial} resulting from mechanical equilibrium conditions on
the trapped unitary gas. According to it, for our system
\begin{equation}
 E_U/2= 2N\langle m\omega^2 R^2/2\rangle .\label{virialr}
\end{equation}

In Table 1 we show the energy per particle at unitarity $E_U/2N$
obtained from the VQMC calculation and the trap energy
$m\omega^2\langle R^2\rangle$ for closed shell configurations with
${\cal M}_F\le 9$.
 It can be noticed
that, within error bars, the virial relation Eq.(\ref{virialr}) is
satisfied. The approximate linear behavior of  $E_U/2N$ as a
function of ${\cal M}_F$:
\begin{equation}
E_U/2N\hbar\omega \sim 0.53\pm 0.01({\cal M}_F + 1.95\pm 0.06)
\end{equation}
 supports the universality relationship Eq.~(\ref{beta}) with the universal
 parameter $\beta=-0.50^{+0.02}_{-0.04}$.

Previous theoretical calculations predict $\beta\sim -0.326$
\cite{heiselberg}, $\beta\sim -0.4$ \cite{baker,engelbrecht},
$\beta\sim -0.56$ \cite{carlson,giorgini}, $\beta =-0.75$
\cite{Dlee}, $\beta= -0.492$ \cite{javier} and  $\beta =-0.545 $
\cite{peralib}. The first experimental measurements yielded
$\beta\sim -0.3$ \cite{salomon}, $\beta=-0.49\pm 0.04$
\cite{kinast}, $\beta=-0.64\pm 0.15$
\cite{bourdel} and $\beta =-0.68^{+0.13}_{
-0.1}$\cite{bartenstein}. More recently the value $\beta
=-0.54\pm0.05$\cite{partridge} and $\beta=-0.54^{+0.05}_{-0.12}$
\cite{jin07} was found for $^6$Li and $^{40}$ K respectively.

The single-particle and the two-particle correlation functions
were calculated as a function of the scattering length. First, we
analyze the density profile as a function of the distance to the
center of the harmonic trap (Fig.~2). The numerical profile of the
ideal gas reproduces satisfactorily the Thomas-Fermi (TF)
prediction. At unitarity, the density resembles more a TF than a
Gaussian profile, as predicted in Ref. \cite{stajic}. For the BEC
regime a clear molecular bunching is observed around the origin,
however, due to molecular repulsion, these bosonic molecules are
prevented from collapsing to the center of the trap.

By analyzing the two point correlation functions, Pauli blocking
can be observed. As expected, the radii at which no atoms of the
same species can be found diminishes as the intensity of the
interaction increases for a given range of the potential
\cite{becbcs}.
 In Fig.~3 we compare the two point correlation
functions
 $K(\vec r_i^\uparrow,\vec r_j^\downarrow)$ of atoms in different hyperfine
states in the ideal and BEC regimes. The enhancement of
correlation for short $r_{ij}$ near the center of the trap
($R_{cm}<0.65\sqrt{\hbar/m\omega}$) indicates molecule formation,
while the increase of probability of finding pairs of particles
separated at relative distances of the order of 1.3 and 2.3 $
\sqrt{\hbar/m\omega}$ with center of mass radii 0.65 and 1.09
$\sqrt{\hbar/m\omega}$ is a manifestation of molecular
condensation. A similar analysis performed for the BCS regime
shows a slight increase of long distance correlations of atoms in
different states.

 Summarizing, we have performed for the first time $direct$
  tests of the universality hypothesis in the unitarity limit for an
 $inhomogeneous$ interacting Fermi gas using VQMC techniques. These tests
 include:
 the $N-$independent energy curve features as described in Fig.~1,
 the verification of virial relations for each $N$ and the variational
 evaluation of $\beta$ using a linear fit of the energy per particle
 $E_U/2N$ as a function of the
 Fermi number ${\cal M}_F$. We have also shown that the optimized
wave functions yield  reasonable values compared with experimental
observations not just for
 the energy per particle,
 but also for the mean radii at the different regimes, the densities
 and the two point correlation functions.

{\bf Acknowledgments}.This work was partially supported by Conacyt
41048-A1 and DGAPA-UNAM contract PAPIIT IN117406-2.

\begin{table}[htdp]
\begin{tabular}{|c|c|c|c|c|c|c|c|c|c|}
\hline
${\cal M}_F$&$N$&$\lambda_{opt}$&{\it\small E}$_{ IFG}${\small /2{\it N}}& {\it \small E}$_U${\small /2{\it N}} & $\beta$&$m\omega^2\langle R^2\rangle$\\
$\quad$&$\quad$&$(\pm 0.005)$&$[\hbar\omega]$&$[\hbar\omega]$& ($\pm$ 0.02)&$[\hbar\omega]$\\
\hline
0&1  & 1   & 1.5 &1  &         -0.5556     &0.99$\pm$0.07\\
1&4  &0.666& 2.25&1.32$\pm$0.05&-0.66&1.44$\pm$0.11\\
2&10 &0.472& 3   &2.00$\pm$0.05&-0.56&1.93$\pm$0.16\\
3&20 &0.382& 3.75&2.48$\pm$0.05&-0.56&2.55$\pm$0.10\\
4&35 &0.333& 4.5 &3.13$\pm$0.05&-0.52&3.01$\pm$0.12\\
5&56 &0.261& 5.25&3.62$\pm$0.05&-0.52&3.57$\pm$0.14\\
6&84 &0.242& 6   &4.25$\pm$ 0.1&-0.50&4.23$\pm$0.16\\
7&120&0.187& 6.75&4.85$\pm$ 0.1&-0.48&4.77$\pm$0.18\\
8&165&0.186& 7.5 &5.25$\pm$ 0.1&-0.51&5.32$\pm$0.20\\
9&220&0.168& 8.25&5.82$\pm$ 0.1&-0.50&5.88$\pm$0.22\\
\hline

\end{tabular}
\caption{Optimized  variational parameter $\lambda$, variational
values of the energy per particle $E_U/2N$ and  the trap energy
$m\omega\langle R^2\rangle$ for closed shells as a function of
${\cal M}_F$ at unitarity. The error bars in the last case take
into account statistical sources and uncertainties in the value of
$\lambda$. The interaction potential range $b/2$ is varied so that
$k_Fb/2= 0.01$. }
\end{table}

\newpage

\begin{figure}
\includegraphics{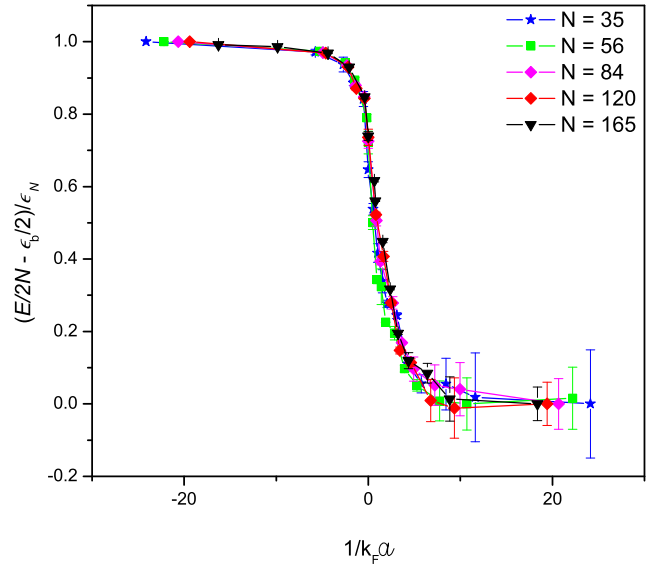}
\caption{  Variational energy per particle $E/2N$ as a function of
$1/k_Fa$.  The normalization factor $\epsilon_N$ depends on the
number of particles and is defined as follows. The ideal gas
energy per particle, $E_{IFG}/2N$, is the asymptotic value of the
energy per particle for small negative scattering lengths
$E(a_-)/2N$; the asymptotic value $E(a_+)/2N$ corresponds to the
energy per particle for small positive scattering lengths (here
$a_+ = 0.012\sqrt{\hbar/m\omega}$) minus half the bound molecular
energy $\tilde\epsilon_b(a_+)/2$ for this value of $a$ (here
$\tilde\epsilon_b(a_+)= 2540.22\hbar\omega$). The $N$ dependent
normalization factor $\epsilon_N$ is the difference between
$E(a_-)$ and $E(a_+)$. Finally $\epsilon_b(a)/2 = \tilde
\epsilon_b(a)/2 + E(a_+)/2N$. In these calculations the range of
the interactions is $b/2 = 0.015 \sqrt{\hbar /m\omega}$. }
\end{figure}
\begin{figure}
\includegraphics{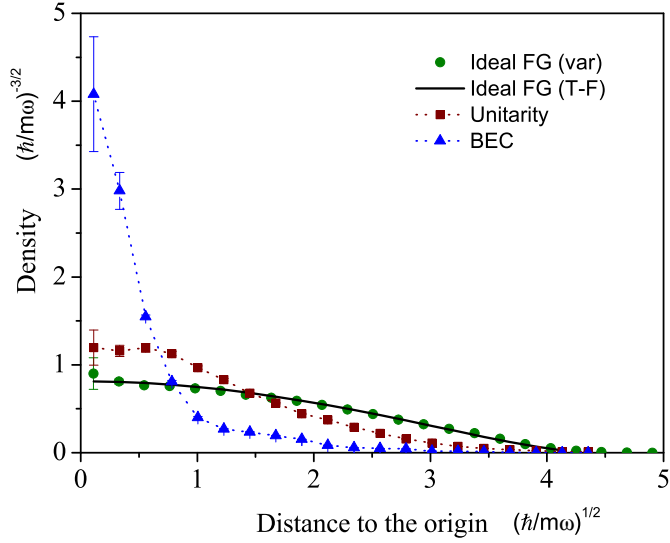}
\caption{ Density profiles for $2N=330$ atoms. The range of the
potential is $b/2 =0.015\sqrt{\hbar/m\omega}$ and the scattering
length for the BEC density corresponds to $1/k_Fa= 3.25$.}
\end{figure}
\begin{figure}
\includegraphics{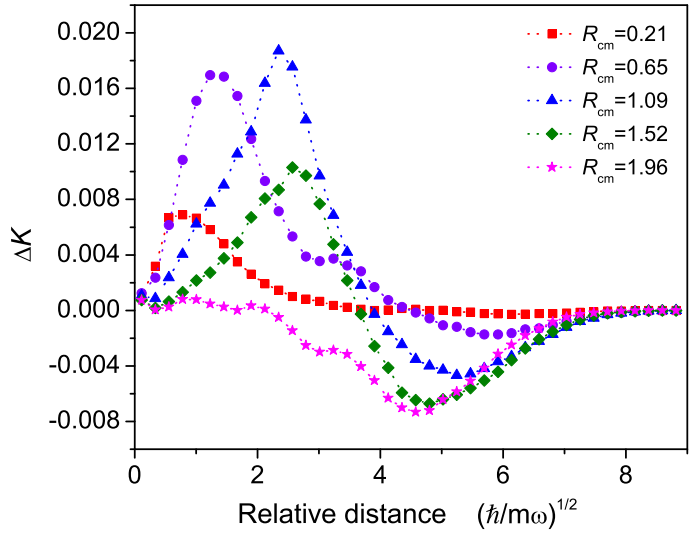}
\caption{Probability difference $\Delta K$ that two particles with
antiparallel spin are found separated a distance $r_{ij}$ in the
BEC and ideal regimes. Each curve in this figure correspond to
different spherical radii $R_{cm}$ measured from the center of the
trap. Calculations are performed at
$b/2=0.015\sqrt{\hbar/m\omega}$ and $1/k_Fa=3.25$. }
\end{figure}
\end{document}